\documentclass[3p,times,procedia]{elsarticle}
\usepackage{ecrc}

\usepackage[linesnumbered,algoruled]{algorithm2e}
\SetKwInput{KwData}{Input} 
\SetKwInput{KwHyperparameters}{Hyperparameters} 
\SetKwInput{KwResult}{Output} 

\usepackage[normalem]{ulem} 
\usepackage{lipsum} 
\usepackage{multicol} 

\usepackage{amsmath}
\usepackage{amssymb}
\usepackage{mathtools}

\usepackage{graphicx} 
\usepackage{subcaption}
\usepackage{multirow} 

\newcommand\vGroup[2]{\underset{#2}{\underbrace{#1} } }

\newcommand{\keepcomment}{1} 
\ifnum\keepcomment=1
	\usepackage[colorinlistoftodos,textsize=scriptsize]{todonotes} 
    \newcommand{\stkout}[1]{\ifmmode\text{\sout{\ensuremath{#1}}}\else\sout{#1}\fi}
    
\else
	
	\usepackage[disable]{todonotes} 
\fi

\usepackage{array} 

\usepackage{enumitem}

\usepackage{titlesec}
\titlespacing\section{0pt}{12pt plus 4pt minus 2pt}{0pt plus 2pt minus 2pt}
\titlespacing\subsection{0pt}{8pt plus 4pt minus 2pt}{0pt plus 2pt minus 2pt}

\volume{00}

\firstpage{1}

\journalname{Transportation Research Procedia}

\runauth{X. Wu et al.}

\jid{trpro}

\usepackage{amssymb}

\biboptions{authoryear}

\usepackage[figuresright]{rotating}

\usepackage[bookmarks=false]{hyperref}
    \hypersetup{colorlinks,
      linkcolor=blue,
      citecolor=blue,
      urlcolor=blue}

\begin{document}
\pagenumbering{gobble}
\begin{frontmatter}



\dochead{26th Euro Working Group on Transportation Meeting (EWGT 2024)}%

\title{Joint Design of Conventional Public Transport Network and Mobility on Demand}


\author[a]{Xiaoyi Wu\corref{cor1}} 
\author[b]{Nisrine Mouhrim}
\author[b]{Andrea Araldo}
\author[c]{Yves Molenbruch}
\author[d]{Dominique Feillet}
\author[e]{Kris Braekers}

\address[a]{Department of Management, Technical Univerisity of Denmark}
\address[b]{SAMOVAR, Télécom SudParis, Institut Polytechnique de Paris}
\address[c]{Vrije Universiteit Brussel, Business Technology and Operations}
\address[d]{Mines Saint-Etienne, Univ. Clermont Auvergne, CNRS, UMR 6158 LIMOS, Centre CMP}
\address[e]{Hasselt University, Faculty of Business Economics}

\begin{abstract}
Conventional Public Transport (PT) is based on fixed lines, running with routes and schedules determined a-priori. In low-demand areas, conventional PT is inefficient. Therein, Mobility on Demand (MoD) could serve users more efficiently and with an improved quality of service (QoS). The idea of integrating MoD into PT is therefore abundantly discussed by researchers and practitioners, mainly in the form of adding MoD \emph{on top of PT}. Efficiency can be instead gained if also conventional PT lines are redesigned after integrating MoD in the first or last mile. In this paper we focus on this re-design problem.  We devise a bilevel optimization problem where, given a certain initial design, the upper level determines stop selection and frequency settings, while the lower level routes a fleet of MoD vehicles. We propose a solution method based on Particle Swarm Optimization (PSO) for the upper level, while we adopt Large Neighborhood Search (LNS) in the lower level. Our solution method is computationally efficient and we test it in simulations with up to 10k travel requests. Results show important operational cost savings obtained via appropriately reducing the conventional PT coverage after integrating MoD, while preserving QoS.
\end{abstract}

\begin{keyword}
Ride-sharing, Mobility-on-demand, Routing Algorithms, Public Transportation, Multi-modal Routes, Mobility as a Service (MaaS), Transport Network Design




\end{keyword}
\cortext[cor1]{Corresponding author. Tel.: +0-000-000-0000 ; fax: +0-000-000-0000.}
 \end{frontmatter}

\email{author@institute.xxx}




\section{Introduction}

It is well known that conventional Public Transport (PT) is inadequate in the suburbs. The sparse demand density in such areas forces PT operators to provide a low-frequency low-coverage service, to prevent the operational cost per passenger to explode. This leads to poor QoS and a chronic car-dependency of the suburban population. 
\cite{Duo2024} assumes that MoD is integrated with conventional PT and a single trip can be composed of conventional PT legs and MoD legs. However, conventional PT is still assumed unchanged, even after MoD is integrated. In \cite{Calabro2023}, the joint design of conventional PT and MoD is proposed, at a strategic level, consisting in deciding in which regions MoD should operate and how conventional PT lines should change. The description of such design remains however a very high level, resorting to approximated density functions and geometrical abstractions.

This paper instead focuses on the tactical and operational aspects of multimodal PT (conventional PT + MoD). We in particular focus on redesigning conventional PT lines in  order to more efficiently exploit the integration with MoD. 
To this aim, we present a bilevel optimization problem. In the \textbf{upper level}, we decide stop activation status and frequencies of conventional PT lines.  In the \textbf{lower level}, we solve instead an Integrated Dial-A-Ride Problem (IDARP) (\cite{Posada2017}), in which MoD fleet sizing and routing decisions are taken, together with user trips. MoD vehicle routes are constructed so as to allow multimodal user trips, composed of conventional PT legs and MoD legs. We solve the upper level via Particle Swarm Optimization (PSO) metaheuristic and we use the Large Neighborhood Search (LNS) metaheurstic from \cite{Mol2020} for solving the lower level.
Table~\ref{tab:related-work} summarizes the related work closest to ours. The main novelty of our work is that we jointly redesign conventional PT lines (via stop selection) and at the same time compute exact routing of MoD vehicles to harmonize with conventional PT lines.
We simulate our solution in a scenario representing in a simplified fashion the Paris region in a low demand period. We concentrate on this period, because that is when MoD becomes more adapted to the demand. Therefore, the output of our method illustrates how multimodal PT should operate in the considered time period and we assume that such structure should change over the day, according to the demand (leaving a larger and larger role to conventional PT during peak hours).
We show that operational cost savings can be achieved if partially (and appropriately) replacing conventional PT with MoD, while still appropriately satisfying the considered demand. Overall, the designs we obtain imply introducing an appropriately sized MoD fleet and reducing the extent of conventional PT, by skipping certain stops (from 27\% to 67\% of stops, depending on the line) and reducing the bus frequency (from 64\% to 94\%, depending on the line) and also completely deactivating certain lines.




\begin{table}[htbp]
    \caption{Closest related work and illustration of the multimodal PT}
    \centering
    \begin{scriptsize}
    \begin{tabular}{p{2cm} p{0.45cm} p{0.45cm} p{0.45cm} p{0.85cm} p{0.35cm} p{0.45cm} p{0.45cm} p{0.45cm} p{7cm}}
    \hline
    Properties & \cite{stiglic2018enhancing} & \cite{Sun2018} & \cite{Pos2017} & \cite{Mol2020} & \cite{lee2017dynamics} & \cite{steiner2020strategic} & \textbf{Our paper} & \cite{Chow2019} & \\
    \hline
    Public transit & Yes & Yes & Yes & Yes\footnote{Just for ambulant persons, whereas wheelchair riders always use a door-to-door service} & Yes & Yes & Yes & Yes&  \multirow{11}{*}{\includegraphics[width=7cm, keepaspectratio]{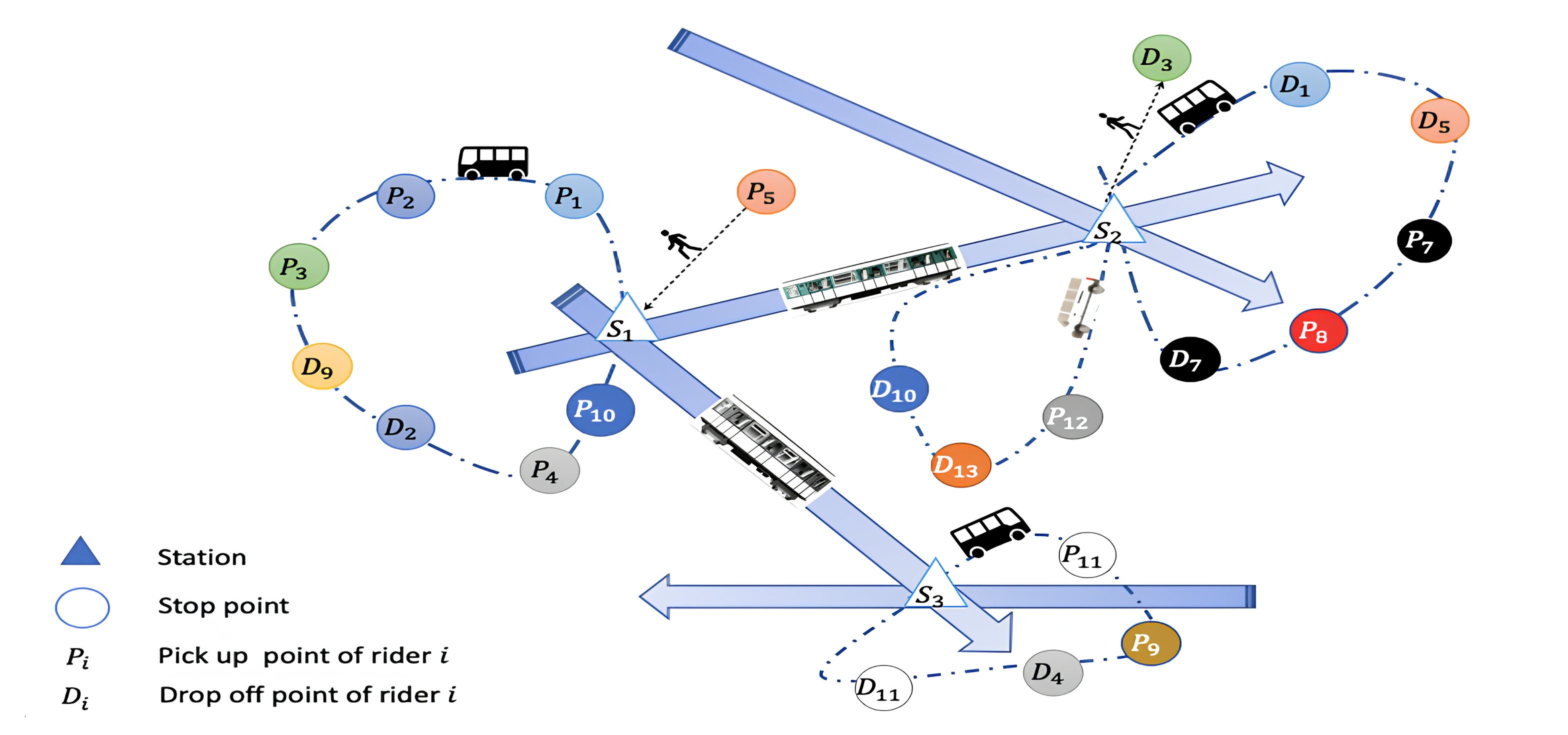}} \\ 
    Online or Offline algorithm (On/Off)& Off & Off & Off & Off & Off & Off & On & On&\\ 
    First mile & Yes & Yes & Yes & Yes & Yes & Yes & Yes & Yes&\\ 
    Last mile & No & No & Yes & Yes & No & Yes & Yes & Yes&\\ 
    Time window constraint & Yes & Yes & Yes & Yes & Yes & No & Yes & No&\\ 
    Maximum ride duration constraint & Yes & Yes & Yes & Yes & No & Yes & No & No&\\ 
    Meeting points & No & Yes & Yes & No & No & & No &&\\ 
    Walking possibility & Yes & No & Yes & No & No & Yes & Yes & Yes&\\ 
    Door-to-door possibility & Yes & No & No & Yes & No & Yes & Yes & Yes&\\ 
    Single or multiple transit line (S/M)& S & S & M & M & S & & M & M&\\ 
    Zones & No & No & No & No & Yes & Yes & No & Yes&\\ 
    Decision variables: Pax (passenger trajectory), Veh (vehicle trajectory)& Veh & Veh & Pax+ Veh& Pax+ Veh & Pax+ Veh & Pax & Pax+ Veh & Pax+ Veh&\\ 
    Heuristic (H) or Exact (E) resolution   & H & H & E & H & E+H & E & H & E+H &\\ 
    \hline
    \end{tabular}
    \end{scriptsize}
    \label{tab:related-work}
\end{table}

\section{System model}
\label{main}

The form of MoD we consider is Ride Sharing (RS).
The objective of the upper level is to decide stop activation status and line frequencies to minimize a cost function, which depends on the number of RS cars and PT vehicles employed. In the lower level, the objective is to minimize the total kilometers traveled by RS cars. Therefore, RS cars can transport users from their origin to their destination directly, or from their origin to a PT stop at which users can board a PT vehicle, or from a PT stop to their final destination. Due to this possibility of transferring between RS and PT, any change in the upper level of the PT layout impacts RS routing in the lower level. On the other hand, the decisions in the lower level result in a certain fleet size, which contributes to the total cost that the higher level aims to minimize.

\subsection{Graph Representation of Conventional Public Transport}
\label{sec:graph}
Public Transport (PT) is defined by set $\mathcal{PS}$ of potential stops and set $\mathcal{L}$ of lines. Each line $l\in\mathcal{L}$ is characterized by a sequence of potential stops $\mathcal{P}_l\subseteq\mathcal{P}$. A decision variable of our optimization problem will determine subset $\mathcal{S}_l\subseteq\mathcal{P}_l$ of stops that are active (stops in $\mathcal{P}_l\setminus\mathcal{S}_l$ will be skipped). \emph{PT graph} $\mathcal{G}$ is composed of active nodes $\mathcal S = \bigcup_{l\in\mathcal{L}} \mathcal{S}_l$ and arcs $(u,v)$, where $u$ and $v$ are consecutive active stops on the same line. We define ${A}$ as the set of arcs, related to all the lines. Any line $l$ serves the sequence of stops $\mathcal{S}_l$ in forward and backward directions and has a frequency~$f_l$, which is a decision variable.
We consider a given time period during which PT graph~$\mathcal G$ remains unchanged. For simplicity, we assume PT vehicles have enough capacity to serve all the demand, as in~\cite[\S3]{Chow2019}. 

Average in-vehicle travel time $t^{TT}_{uv}$ on an arc $(u,v) \in \mathcal{A}$ is known and independent of the line. We compute it as $t^{TT}_{uv} = d(u,v)/\nu_\text{PT}$,  where $d(u,v)$ is the Euclidean distance between stops $u$ and and $v$ and $\nu_\text{PT}$ is the average speed of a PT vehicle.
Let $t^{PT}_u$ be the dwell time at a stop $u \in \mathcal S$, i.e., the time a PT vehicle stays in a stop for passenger boarding and alighting. We compute average time $t^l_{uv}$ needed to go from stop $u\in \mathcal{S}_l$ to stop $v\in \mathcal{S}_l$ (not necessarily consecutive) along a single line $l$:
\vspace{-1cm}
\begin{equation}
    t^l_{uv} = 
    \underbrace{
    \frac{1}{2 f_l}}_{
    \substack{
        \text{waiting time}
        \\
        \text{\cite[(2.4.28)]{Cascetta2009}}
    }
    } 
    + \underbrace{
        \sum_{
            \substack{i,j \in \mathcal{S}_l(u,v) 
            \\ i,j\text{ consecutive}}} t^{PT}_{ij}}_{
            \text{in-moving vehicle travel time}
    }
    + \underbrace{\sum_{i \in \mathcal S_l(u,v)\setminus\{u\}} t^{PT}_i}_{\text{dwell time}}
\label{eq:tluv}
\end{equation}
\vspace{-0.7cm}

\noindent where $S_l(u,v)$ indicates the sequence of active stops between stop $u$ and $v$. To compute trips between any two stops of a certain PT graph $G = (\mathcal{S},\mathcal{A})$, we need to define an additional graph 
$\mathcal{G}'=(\mathcal{S},\mathcal{A}')$, constructed as follows. There is an arc $(u,v)^l$ in $\mathcal{A}'$, whenever there exists a line $l$, such that is possible to go from $u$ to $v$ along that line. The time associated to this arc is~\eqref{eq:tluv}.
In case multiple lines connect the same pair of stops, there is a connecting arc $(u,v)^l\in\mathcal A'$ for each line~$l$. Therefore, $\mathcal{G}'$ is a multi-graph.
%
Using graph $\mathcal{G}'$, the minimal average traveling and waiting time  $t_{uv}$ between two active stops $u,v \in \mathcal{S}$ can be computed by solving a shortest path problem in multigraph $\mathcal{G}'$. Let us consider a path $\mathcal{P}=[(u_1,v_1)^{l_1},\dots,(u_m,v_m)^{l_m}]$ of $\mathcal{G}'$, where $v_i=u_{i+1}$. Such path means entering PT via stop $u_1$, boarding line $l_1$ and going up to stop $v_1$, changing for line $l_2$ to go from stop $v_1=u_2$ to stop $v_2$ etc. Denoting with $t_\text{ingress}$ and $t_\text{egress}$ the time to go from the road to the PT stop and vice-versa, and denoting with $t_\text{change}$ the time to walk when changing from a line to another, the travel time is of such path is
\vspace{-1cm}\begin{align}
    t_\mathcal{P} =
    t_\text{ingress} 
    + \vGroup{\sum_{i=1}^m t_{u_i,v_i}^{l_i} }{\text{within-lines}}
    + (m-1) \cdot t_\text{change}
    + t_\text{egress}.
    \label{eq:tP}
\end{align}
\vspace{-0.8cm}

For any pair of stops $u,v\in \mathcal{S}$, $\mathcal{P}^*(u,v)$ is the shortest path, i.e., the path that minimizes quantity~\eqref{eq:tP}. We assume that, whenever a user enters stop $u$ at instant $t$, she will reach stop $v$ at instant $t+t_{\mathcal{P}^*(u,v)}$. 

\subsection{User trips} 
\label{sec:user-demand}
We represent $n$ users as a set of nodes (note that they are not the nodes of PT graph $\mathcal{G}$) corresponding to their origin, destination and transfer nodes, detailed as follows.
Set $\mathcal{N}$ consists of an artificial depot node $0$, a set of origin nodes $\mathcal{O} = \{1,..., n\}$, a set of destination nodes $\mathcal{D} = \{n+1,..., 2n\}$. With slight abuse of notation, we will represent by $i$ the origin of a user and the user itself, and $i+n$ referring to the corresponding destination. We also define set $\mathcal T_i$ of potential transfer nodes at which user~$i$ can switch from Ride Sharing (RS) to PT (or vice-versa). To define~$\mathcal T_i$, we duplicate set~$\mathcal S$ of active stops per each user: $
\mathcal{T}_i =
2n+(i-1)\cdot|\mathcal S|+1,\,\,\,\,2n+(i-1)\cdot|\mathcal S|+2,\,\,\,\,\dots,2n+(i-1)\cdot|\mathcal S|+|\mathcal S|$. The entire set of possible transfer nodes is~$\mathcal T= \bigcup_i \mathcal T_i$ and~$\mathcal N = \mathcal O \cup \mathcal D \cup \mathcal T$.

Let~$\mathcal{R}$ be the set of all trip requests. 
We partition it into
$
\mathcal{R}=
\mathcal{R}_W\cup 
\mathcal{R}_\text{PT}\cup 
\mathcal{R}_\text{RS}\cup 
\mathcal{R}_\text{W-PT-RS} \cup 
\mathcal{R}_\text{RS-PT-W}.
\label{eq:partition}
$, i.e., requests of users who will just walk, or will be served by PT only, or will be served by RS only, or will walk in the first mile, enter PT in a certain stop and then use RS in the last mile, or will use RS in the first mile, then enter PT to a certain stop and finally walk in the last mile, respectively. 
For simplicity, we do not consider trips which have RS in both first and last mile. This assumption can be later removed, following an approach similar to~\cite{Chow2019}.

The procedure to partition users is described in Alg.~\ref{alg:partition}. More complex mode choice models are out of scope here and could be integrated later. In broad terms, our partitioning assume that users would walk if possible, otherwise use PT if possible, otherwise use PT+RS if possible, otherwise use RS. It is reasonable to assume these preferences, as no monetary cost is associated with walking, some monetary cost is associated with PT and larger monetary cost is associated with RS, so that a user would use RS only if no other feasible alternatives are available. In our case ``if possible'' means that the required walking time is less than a certain $d_\text{walk}^\text{max}$ and the trip duration is less than a maximum duration $M_i$ tolerated by user $i$. We assume that $M_i$ is such that at least a direct trip via RS is shorter than $M_i$.

\begin{algorithm}[]
\begin{multicols}{2}
\begin{scriptsize}
\SetAlgoLined
\KwData{
    Set $\mathcal{R}$ of users;
    Maximum walking distance $d_\text{walk}^\text{max}$; 
    Set $\mathcal{S}^k_i$, $\forall$ origin or destination $i$: the $k$ closest stops to $i$; 
    PT graph $\mathcal{G}$
}
\KwHyperparameters{
    $M_i$: maximum trip time user $i$ can tolerate; 
    $\tau_\text{RS}$: threshold on RS feeder travel duration;
    $d_\text{walk}^\text{max}$
}
\KwResult{$\mathcal{R}',\mathcal{R}_W,\mathcal{R}_{PT},\mathcal{R}_\text{RS},\mathcal{R}_\text{W-PT-RS},\mathcal{R}_\text{RS-W-PT}$}
\textbf{Initialize}
$\mathcal{R}=\mathcal{R}'=\mathcal{R}_W=\mathcal{R}_{PT}=\mathcal{R}_\text{RS}=\mathcal{R}_\text{W-PT-RS}=\mathcal{R}_\text{RS-W-PT}=\emptyset$ 
\\
\For{User $i\in\mathcal{R}$}
{
    Take origin $i$ and destination $i+n$
    \\
    \eIf{$d(i,i+n)\le d_\text{walk}^\text{max}$}
    {
        $\mathcal{R}_W = \mathcal{R}_W\cup\{i\}$
    }
    {
        \eIf{$\exists$ active stops $s,s'\in\mathcal{S}: d(i,s)+d(s', i+n)\le d_\text{walk}^\text{max}$}
        {
            Compute shortest path $\mathcal{P}^*(s,s')$ within PT network\\
            \eIf{$t_\text{walk}(i,s)+t_{\mathcal{P}^*(s,s')}+t_\text{walk}(s',i+n)\le M_i$}
            {
                $\mathcal{R}_\text{PT} = \mathcal{R}_\text{PT}\cup\{i\}$
            }
            {
                $\mathcal{R}'=\mathcal{R}'\cup\{i\}$    
            }
        }
        {
            $\mathcal{R}'=\mathcal{R}'\cup \{i\}$
        }
    }
}
// We have now partitioned $\mathcal{R}=\mathcal{R}'\cup\mathcal{R}_\text{PT}\cup \mathcal{R}_\text{W}$\\
// In the next lines we are going to partition $\mathcal{R}'$\\
\For{User $i\in\mathcal{R}'$}
{
    Take origin $i$ and destination $i+n$\\
    \eIf{For any pair of active stops $s,s'\in\mathcal{S}$, we have $d(i,s)>d_\text{walk}^\text{max}$ and $d(s',i+n)>d_\text{walk}^\text{max}$}
    {
        // User $i$ cannot reach any close stop, neither in the first nor the last mile. User $i$ will need a door-to-door RS trip
        \\
        $\mathcal{R}_\text{RS}=\mathcal{R}_\text{RS}\cup\{i\}$
    }
    {
        \uIf{$\exists s,s'\in\mathcal{S}$ s.t. $d(i,s)\le d_\text{walk}^\text{max}$ \textbf{and} $t_\text{walk}(i,s)+t_{\mathcal{P}^*(s,s')}+\tau_\text{RS}\le M_i$}
        {
            $\mathcal{R}_\text{W-PT-RS}=\mathcal{R}_\text{W-PT-RS} \cup \{i\}$
        }
        \uElseIf{$\exists s,s'\in\mathcal{S}$ s.t. $d(s',i+n)\le d_\text{walk}^\text{max}$ \textbf{and} 
        $\tau_\text{RS}+t_{\mathcal{P}^*(s,s')}+t_\text{walk}(s',i+n)\le M_i$}
        {
            $\mathcal{R}_\text{RS-PT-W}=\mathcal{R}_\text{RS-PT-W} \cup \{i\}$
        }
        \uElse{
            $\mathcal{R}_\text{RS} = \mathcal{R}_\text{RS} \cup \{i\}$
        }
    }
}
\end{scriptsize}
 \caption{Partition of users}
 \label{alg:partition}
\end{multicols}
\end{algorithm}

\subsection{Ride Sharing}
\label{sec:mod}

The Mobility in Demand service we consider is Ride Sharing (RS): a fleet of cars can pickup and dropoff passengers. we use the model of RS and the calculation of routing for RS cars from~\cite{Mol2020}.

In order to compute the time window that RS needs to ensure, we have to consider some constraints related to the quality of service demanded by users.
Let $y$ be any location (it can be a stop or not) and suppose we want to ensure a user $i$ arrives at $y$ no later than instant $l_i$. We can compute the latest time at which a user can depart from a stop $s$ to arrive at location $y$ no later than instant $l_i$ as 
\vspace{-0.8cm}\begin{align}
    \label{eq:latest-departure}
    t_\text{max}(s,y,l_i)
    &=
    \sup\left\{ 
    t | 
    t+ t_{\mathcal{P}^*(s,v) } +t_\text{walk}(v,y) \le l_i, 
    \forall v\in \mathcal{S}  
    \right\} 
    =
    \sup\left\{ 
    t = l_i - t_{\mathcal{P}^*(s,v) } - t_\text{walk}(v,y) |
    \forall v\in \mathcal{S}  
    \right\}
\end{align}
\vspace{-1cm}

Similarly, we focus now on a user staying at location $y$ and willing to reach stop $s$ as early as possible and willing to depart from $y$ no earlier than instant $e_i$. The earliest arrival time at $s$ is:
\vspace{-0.7cm}\begin{align}
    \label{eq:earliest-arrival}
    t_\text{min}(y,s,e_i) = 
    \inf \left\{ 
        t=e_i + t_\text{walk}(y,v)) + t_{\mathcal{P}^*(v,s)}
        | v\in \mathcal{S}
    \right\}
\end{align}
\vspace{-1cm}

Let set $\mathcal{R}'= \mathcal{R}_\text{RS}\cup \mathcal{R}_\text{W-PT-RS} \cup \mathcal{R}_\text{RS-PT-W}$ contain all requests that must be handled by RS, either entirely or partially. By construction (Alg.~\ref{alg:partition}), in absence of RS, such users would need to walk more than $d_\text{walk}^\text{max}$ (summing the first and last mile walk) or would take longer than the total tolerated trip time $M_i$.  We assume all requests in $\mathcal{R}'$ need to be served and that the demand is inelastic, i.e., $\mathcal{R}$ does not change when changing the PT network layout or RS routing. However, $\mathcal{R}'$ does change every time we change the PT network layout, as a result of running Alg.~\ref{alg:partition} with a new PT graph $\mathcal{G}$. 

Every user $i$ is associated to an origin node $i$ and corresponding destination node $i+n$. RS users will need appropriate time constraints to be respected. To calculate such constraints, we treat $\mathcal{R}_\text{RS}, \mathcal{R}_\text{W-PT-RS}$ and $\mathcal{R}_\text{RS-PT-W}$ differently:

\begin{itemize}[noitemsep,topsep=0pt,parsep=0pt,partopsep=0pt]
\item For every user $i\in \mathcal{R}_\text{RS-PT-W}$, we assume latest arrival time $l_i$ at destination is exogenously determined. We compute the earliest departure time at the origin as $e_i=l_i-M_i$, where $M_i$ is the maximum tolerable trip time of user $i$ (including waiting). 
Let $\mathcal{T}_i \subset \mathcal{T}$ be the set of potential transfer nodes available to user $i$. This set may be a subset of all eligible PT stops, e.g. the $k$ closest stops to the user's origin. If the user is dropped-off by RS in a certain transfer node, then they will traverse a PT path. If a user uses transfer node $s \in \mathcal{T}_i$, RS should drop them at~$s$ at time~$l_s$ such that, departing from~$s$ the user can reach destination~$i+n$ before~$l_i$. Via~\eqref{eq:latest-departure}, the latest possible arrival time $l_s$ at stop $s$ is
    $l_s = t_\text{max}(s, i+n, l_i)$.

\item For every user~$i\in \mathcal{R}_\text{W-PT-RS}$, earliest possible departure time $e_i$ at origin $i$ is assumed to be exogenously determined. The latest arrival time $l_i$ at destination is computed as $e_i+M_i$. For every transfer node $s \in \mathcal{T}_i$, the earliest possible RS departure time $e_s$ is computed via~\eqref{eq:earliest-arrival} such that the user, leaving their origin after~$e_i$, reaches stop~$s$ (via walking and conventional PT) no earlier than~$e_s$, i.e., 
$
    e_s = t_\text{min}(i,s,e_i)
$.
\item Users $i\in \mathcal{R}_\text{RS}$ declare the latest arrival time at destination, such as $t+M_i$, where $t$ is the time instant at which user $i$ submits their trip request.
\end{itemize}

Let us denote $\mathcal{V}$ the set of RS cars, each with capacity $Q$. We assume they all start and end their activity at a depot $0$. For any pair of nodes $i, j$, the travel time by RS is
$
    t_{i,j}= d(i,j)\cdot \textit{circ} / \nu_\text{car}
$,

where $\nu_\text{car}$ is the average speed of a car and $\textit{circ}\ge 1$ is the circuity (\cite{Boeing2019}), which accounts for the fact that a real world topology implies that any movement from $i$ to $j$ is longer than the Euclidean distance. 

\section{Optimization Problem}
\label{sec:optimization}

\subsection{Bilevel optimization problem definition}
\label{sec:bilevel}
We solve a bilevel optimization problem.
In the \textbf{upper level}, we fix the following decision variables:
\begin{itemize}[noitemsep,topsep=0pt,parsep=0pt,partopsep=0pt]
\begin{small}
\item Number $N_l$ of PT vehicles for each line $l$ and, as a consequence, average frequency
$
    f_l = N_l / (2 t_l),
$ \cite[(2.4.28)]{Cascetta2009},
 where $t_l$ is the time for a PT vehicle to  go from the beginning to the end of line $l$ and $f_l$ is the frequency of that line. Consequently, the minimum and mimum vehicle number $N_{min}$ and $N_{max}$ is dependent by the maximum and minimum frequency, which are 0.25 and 0.06 vehicles/min.
\item Set $\mathcal{S}_l\subseteq\mathcal{P}_l$ of stops to activate in each line $l$ (represented as a vector of binary variables indicating whether a stop in~$\mathcal{P}_l$ is activated or not). 
\end{small}
\end{itemize}

A \emph{PT layout} $\textbf{y}$ (or \emph{solution}) is a vector of values for the decision variables above. $\mathcal{G}(\mathbf{y})$ is the resulting PT graph (\S\ref{sec:graph}). Fixing a PT layout $\mathcal{G}(\mathbf{y})$ also induces a certain partition of users, calculated with Alg.~\ref{alg:partition}, indicating whether each user walks, use PT, use RS or a combination of such modes.

The \textbf{lower level} gets set $\mathcal{R}'$ of user using RS either entirely or partially (\S\ref{sec:mod}) as input. The lower level decides:
\begin{itemize}[noitemsep,topsep=0pt,parsep=0pt,partopsep=0pt]
\begin{small}
\item The fleet size $N^\text{RS}=|\mathcal{V}|$ of active cars in the RS service.

\item The route of all ride-sharing cars, i.e., the sequence of pickups and dropoffs

\item The precise trip of each user, i.e., (i)~the instant and the location in which they will be picked up and dropped off (either at the origin, destination or some transfer stop), (ii)~the exact path traveled within conventional PT (if any), including changes from a line to another (if any), (iii)~the trajectory traveled within a RS car (if any), including possible stopovers to serve other passengers, (iv)~the walking legs.
\end{small}
\end{itemize}

In the lower level, decisions are calculated as in~\cite{Mol2020}, with the objective to minimize kilometers traveled by RS cars, subject to serving all users~$\mathcal{R}'$ (users using RS either for the entire trip or for a part of it) and respecting the time constraints specified in \S~\ref{sec:mod}. The resolution method is Large Neighborhood Search (LNS).

To compute the cost of a solution~$\mathbf{y}$, we assume that a PT vehicle has an operating cost that is $\beta$ times the one of a RS car. We wish to minimize the following expression, which is a proxy of the operating cost of the multimodal system composed of conventional PT and a RS service:
\vspace{-0.7cm}\begin{align}
    f(\mathbf{y})=N^\text{RS} + \beta\cdot \sum_{l\in \mathcal{L}} N_l
    \label{eq:cost}
\end{align}
\vspace{-1cm}

\subsection{Particle Swarm Optimization (PSO)}
\label{sec:pso}
We adapt Particle Swarm Optimization (PSO) to solve the upper level (Alg.~\ref{alg:PSO}). A particle $p$ corresponds to a sequence of PT layouts, evolving along epochs. The set of particles is called \emph{swarm}. Saying that a particle evolves means that the corresponding layout changes from~$\mathbf{y}$ in an epoch to~$\mathbf{y}'$ in the following epoch. Such changes are activation/deactivation of some stops and the number of PT and RS vehicles (\S\ref{sec:bilevel}).
For any particle~$p$, in addition to its layout~$\mathbf{y}_p$ at the current epoch, we also keep $\mathbf{y}_p^\text{ibest}$ the best ``version'' of particle $p$ across all previous epochs (the one with the best performance~\eqref{eq:cost}). $\mathbf{y}^\text{gbest}$ is the best particle among the whole swarm and all epochs, up to the current epoch.
The evolution of each particle~$p$ at each epoch is obtained by two perturbations: Binary PSO (BPSO) (Alg.~\ref{alg:BPSO} - inspired by \cite{Kha2007}) activates/deactivates conventional PT stops and Discrete PSO (DPSO) (Alg.~\ref{alg:DPSO} - inspired by \cite{cipriani2020}) changes the number of PT vehicles per line.

\begin{algorithm}[]
\begin{multicols}{2}
\begin{scriptsize}
\SetAlgoLined
\KwData{Initial PT layout $\mathbf{y}^0$}
\KwResult{The best PT layout $\mathbf{y}^\text{best}$}

\textcolor{gray}{\textit{//Initialize the swarm with the initial version of $P$ particles}} \\
\For{particle index $p=1,\dots,P$} 
{
    \textcolor{gray}{\textit{// Generate particle $\mathbf{y}_p$ as follows}}\\
    \For{Every stop $s\in\mathcal{P}$ of every line $l\in\mathcal{L}$}
    {
        Set~$s$ to active with probability 0.5 \\
        Generate number $N_l$ of PT vehicles in $l$, unif. at rnd in~$[N_\text{min}, N_\text{max}]$
    }
    $\mathbf{y}_p^\text{ibest}=\mathbf{y}_p$
}\label{ln:repeat}
$\mathbf{y}^\text{gbest}=\arg\max\limits_{p=0,1,\dots,P} f(\mathbf{y}_p)$\\

\textcolor{gray}{\textit{//Associate a ``velocity'' to each particle, line and stop}}
\\
\For{each particle index $p=1,\dots,P$, line $l\in\mathcal{L}$, stop $s\in\mathcal{P}$}
{
    $v_{p,0}(l,s)=0; v_{p,1}(l,s)=0$
}

\Repeat{number of epochs}
{
    \For{particle index $p=1,\dots,P$}
    {
        Evaluate cost $f(\mathbf{y}_p)$ via~\eqref{eq:cost}. \\
        \textcolor{gray}{//\textit{Compare performance $f(\mathbf{y}_p)$ to the best version particle~$p$}:} \\
          \If{$f(\mathbf{y}_p) < f(\mathbf{y}_p^\text{ibest})$}
        {
              $\mathbf{y}_p^\text{ibest} = \mathbf{y}_p$
        }
        \textcolor{gray}{//\textit{Compare performance $f(\mathbf{y}_p)$ to the globally best particle $\mathbf{y}^\text{gbest}$}:} \\
        \If{$f(\mathbf{y}_p) < f(\mathbf{y}^\text{gbest})$}
        {
              $\mathbf{y}^\text{gbest} = \mathbf{y}_p$
        }
        \textcolor{gray}{\textit{//Perturb the particle}}\\
        $\mathbf{y}_p=\text{BPSO}(p)$
        \textcolor{gray}{\textit{//Alg.~\ref{alg:BPSO}}}
        \\
        $\mathbf{y}_p=\text{DPSO}(p)$
        \textcolor{gray}{\textit{//Alg.~\ref{alg:DPSO}}}

        Compute the number of RS cars via the low level optimization
    }
 }
 \end{scriptsize}
 \caption{\begin{small}Particle Swarm Optimization (PSO) for the upper level problem.\end{small}}
 \label{alg:PSO}
\end{multicols}
\end{algorithm}

\begin{figure}[htbp]
    \centering
    \begin{minipage}{0.48\textwidth}
        \begin{algorithm}[H]
            \begin{scriptsize}
            \SetAlgoLined
            \KwHyperparameters{
                Constants~$C_1, C_2$
            }
             \KwData{ 
                Index $p$ of a particle, \\
                Previous velocities~$v_{p,0}(l,s),v_{p,1}(l,s)$, $\forall$line $l\in\mathcal{L}$ and  stop $s\in\mathcal{P}_l$
             }
            \KwResult{ Updated particle $\mathbf{y}_p$}
             \For{each line $l\in\mathcal{L}$ and stop $s\in\mathcal{P}_l$}
             {
             Generate random values: $r_1, r_2$ uniformly at random in~$[0,1]$ \\
              \eIf{ Stop~$s$ is active in $\mathbf{y}_p^\text{ibest}$}{
                $d^1_1 = C_1\cdot r_1$\\
                $d^1_0 = -C_1\cdot r_1$\\
               }{
               $d^1_0 = C_1 \cdot r_1$\\
               $d^1_1 = -C_1\cdot r_1$\;
              }
              \eIf{Stop~$s$ is active in $\mathbf{y}^\text{gbest}$ }{
                $d^2_1 = C_2\cdot r_2$\\
                $d^2_0 = -C_2\cdot r_2$\\
               }{
               $d^2_0 = C_2\cdot r_2$ \\
               $d^2_1 = -C_2 \cdot r_2$\;
              }
              Generate value $\textit{inertia}$ uniformly at random in $[-1,1]$ \\
              \textcolor{gray}{\textit{//Update velocities:}}
              \\
                $v_{p,1}(l,s) = \textit{inertia}\cdot v_{p,1}(l,s)+d^1_1+d^2_1$ 
                \\
                $v_{p,0}(l,s) = \textit{inertia} \cdot v_{p,0}(l,s)+d^1_0+d^2_0$ 
                \\               
              $v
              =
              \begin{cases}
                  v_{p,1}(l,s) & \text{Stop }s\text{ is active in }\mathbf{y}
                  \\
                  v_{p,0}(l,s) & \text{Otherwise}
              \end{cases}
              $
              \\
              With probability~$\textit{sigmoid}(v)$, activate stop~$s$, else deactivate it
              }
            \end{scriptsize}
             \caption{BPSO for stop (de)activation}
              \label{alg:BPSO}
            \end{algorithm}
    \end{minipage}\hfill
    \begin{minipage}{0.48\textwidth}
        \begin{algorithm}[H]
            \begin{scriptsize}
            \SetAlgoLined
             \KwHyperparameters{Constants~$CR_1, CR_2, CR_3$. By increasing $CR_1$ we tend to guide particles closer to~$\mathbf{y}_p^\text{ibest}$. 
Higher values of~$CR_2$ force particles to resemble~$\mathbf{y}^\text{gbest}$.
$CR_3$ allows to tune the level of randomness of the particles. 
}
             \KwData{ Particle index $p$}
             \KwResult{Modified particle $\mathbf{y}_p$}
             \For{each line $l\in\mathcal{L}$}
             {
                Generate $r$ uniformly at random in $[0,1]$\\
                $ ac = CR_1 \cdot r$ \\
                \eIf{$ ac \le 0.5 $ }
                {
                    $N_l^\text{aux1}=y_z(N_l)$
                }
                {
                    $N_l^\text{aux1}=y_z^\text{ibest}(N_l)$
                }
                Generate another $r$ uniformly at random in $[0,1]$\\
                $ac = CR_2\cdot r$ \\
            \eIf{$ ac \le 0.5 $ }
              {
                  $N_l^\text{aux2}=N_l^\text{aux1}$
              }
              {
                  $N_l^\text{aux2}=y^\text{gbest}(N_l)$
              }

                Generate another $r$ uniformly at random in $[0,1]$\\
    $ac = CR_3 \cdot r$ \\
    \eIf{$ac \le 0.5$ }
  {
      $y_z(N_l)=N_l^\text{aux2}$
   }
   {
    Generate $y_z(N_l)$ uniformly at random in $[N_\text{min}, N_\text{max}]$.
   }
}
\end{scriptsize}
             \caption{DPSO for number of PT}
              \label{alg:DPSO}
            \end{algorithm}
    \end{minipage}
\end{figure}

\section{Simulation results}

\begin{table}[]
    \centering
    \begin{scriptsize}
    \begin{tabular}{c c c }
    \hline
    \multicolumn{3}{c}{\textbf{Hyperparameters of the Discrete Particle Swarm Optimization (DPSO) algorithm}}\\
    \hline
        $CR_1$, $CR_2$, $CR_3$ &  0.55, 0.65, 0.52 (respectively) 
    \\
    \hline
    \multicolumn{3}{c}{\textbf{Evaluation scenario (default values in underlined bold)}}\\
    \hline
    RS Car speed $\nu_\text{car}$ & 
    30 Kmh &
    Normal traffic~\cite{yong2011traffic}
    \\
    PT vehicle speed $\nu_\text{PT}$ &
    60 Kmh &
    Metro line~\cite{dominguez2014multi}
    \\
    
    Walking speed $\nu_\text{walk}$ &
    1.4 m/s (5.04 Kmh) &
    Google Maps
    \\
    
    Car circuity $\text{circ}$ &
    1.255 &
    \cite{Levinson2015}
    \\
    
    Walk circuity $\text{circ}_\text{walk}$ &
    1.391 &
    \cite{zhao2013relationship}
    \\
    
    $t_\text{ingress}, t_\text{egress}$ &
    Both are 0 &
    \\
    
    Max walk distance $d_\text{walk}^\text{max}$ &
    2.52 Km ($\sim$30 min) &

    \\
    Ingress, change and egress times $t_\text{ingress}=t_\text{change}=t_\text{egress}$ (\S\ref{sec:graph})
    & 0 
    \\    
    Dwell time $t_i^{PT}$ of PT vehicle at a stop (includes time for acc(dec)elerating) & 3 minutes  &
    \\
    RS car pickup and dropoff time (includes time for acc(dec)elerating) &  1 minute &
      \\ 
    Minimum bus headway (to avoid bus bunching) & 2 minutes & \cite{AntoniouSadrani2022}
    \\
    Maximum lengthening~$\gamma$ & 
    $\gamma\in\{1, 1.25, \underline{\mathbf{1.5}}, 1.75, 2, 2.5, 3\}$
    &
    \\
    Number of users (i.e., of trips) & 
    $\{100, 500, \underline{\mathbf{1000}}, 5000, 10000\}$
    &
    \\
    Maximum trip time~$M_i$ tolerated by user $i$
    & $M_i = \gamma \times$ direct trip time by a private car
    &
    \\
    Cost of operating a bus
    & $\beta=2\times$ cost of operating a RS car
    & \cite{bosch2018cost}, \cite[\begin{tiny} Tab~3, $\gamma_c$\end{tiny}]{CatsJointOptim2021}
    \\
    Number of epochs of PSO for every scenario
    & 50
    &
    \\
    Processor and RAM of the PC used get our results & Threadripper 3970X, 128GB RAM &
    \\
    Maximum time needed to run a single simulation & 2318.77s (38.65 minutes) &
    \\
    \hline
    \end{tabular}
    \end{scriptsize}
    \caption{Parameters considered}
    \label{tab:parameters}
\end{table}

\begin{table}[htbp]
    \caption{PT layout changes in the default scenario}
    \centering
    \begin{footnotesize}
    \begin{tabular}{c c c c c }
    \hline
    \textbf{Line} & \textbf{Skipped stops}  & \textbf{Reduction of} & \textbf{Num of}  & 
    \multirow{8}{*}{\includegraphics[height=3.2cm, keepaspectratio]{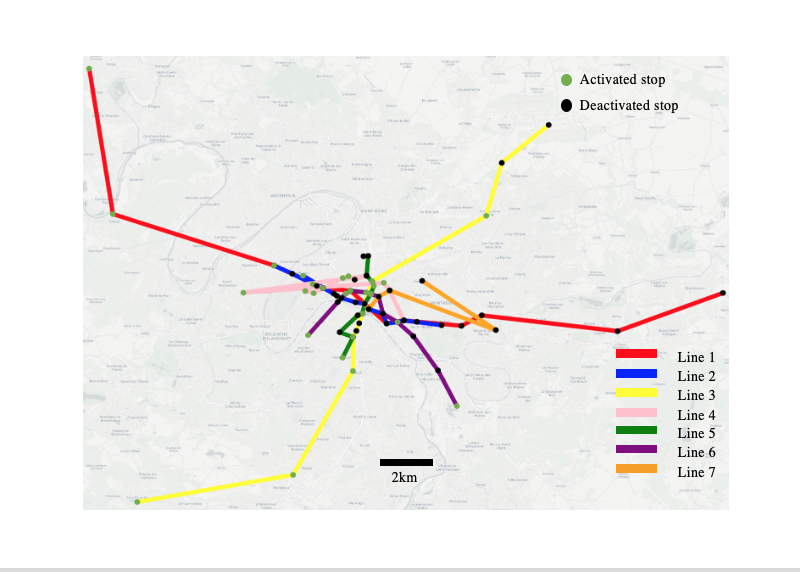}}
    \\
    & \textbf{fraction} &  \textbf{num of buses} & \textbf{users} &\\
    1 & 42\% & 76\% & 105 & \\
    2 & 67\% & 80\% & 15 &  \\
    3 & 50\% & 64\% & 131 & \\
    4 & 27\% & 88\% & 6 & \\
    5 & 41\% & 84\% & 34 &  \\
    6 & 38\% & 94\% & 17 &  \\
    7 & 100\% & 100\% & 0 &  \\
    \hline
    \end{tabular}
    \end{footnotesize}
    \label{tab:PT_layout_under_default_scenarios}
\end{table}




We conduct numerical experiments simulated in an area with the same size of the Paris region, with~$7$ conventional PT lines, approximately corresponding to part of the PT lines in Paris region (see figure inside Table~\ref{tab:PT_layout_under_default_scenarios} and parameters in Table~\ref{tab:parameters}).
%
For simplicity, travel time $t_{ij}^{RS}$ on all arcs ending and starting with the depot are set to 0. 

To simulate the transportation demand, we distribute user requests over a three-hour time window. The spatial distribution of these requests mirrors the geographic characteristics of the Paris region, which is divided into three main zones: Paris (Central Zone), the Inner Suburbs, and the Outer Suburbs ~\cite[page~12]{IdF2019}. We consider inter-zone and intra-zone travel requests to capture the diversity of transportation needs across the region.

Under different scenarios, we compare~$N^0=f(\mathbf{y}^0)$ and~$N^*=f(\mathbf{y}^*)$, i.e., the cost of the initial and optimized solution, respectively. In all scenarios, solution~$\mathbf{y}^0$ is initialized with~$\sum_{l\in\mathcal{L}}N_l=25$ buses ($N_\text{RS}$ is then determined in the lower level - \S\ref{sec:bilevel}). For every scenario, to obtain optimized solution~$\mathbf{y}^*$, we start from~$\mathbf{y}^0$ and we first perform our optimization (\S\ref{sec:pso}) setting maximum lengthening parameter~$\gamma=1$ (Table~\ref{tab:parameters}) to obtain~$\mathbf{y}^*$ and the corresponding cost~$N_{\gamma=1}^*$. Then, for~$\gamma=1.25$, we start the optimization with the previously found~$\mathbf{y}^*$ and we perform our optimization to obtain a new optimal solution~$\mathbf{y}^*$ and the corresponding cost~$N_{\gamma=1.25}^*$.  We continue up to $\gamma=3$.

\begin{figure}[htbp]
    \centering
    \begin{subfigure}{.32\textwidth}
        \centering
        \includegraphics[width=\linewidth]{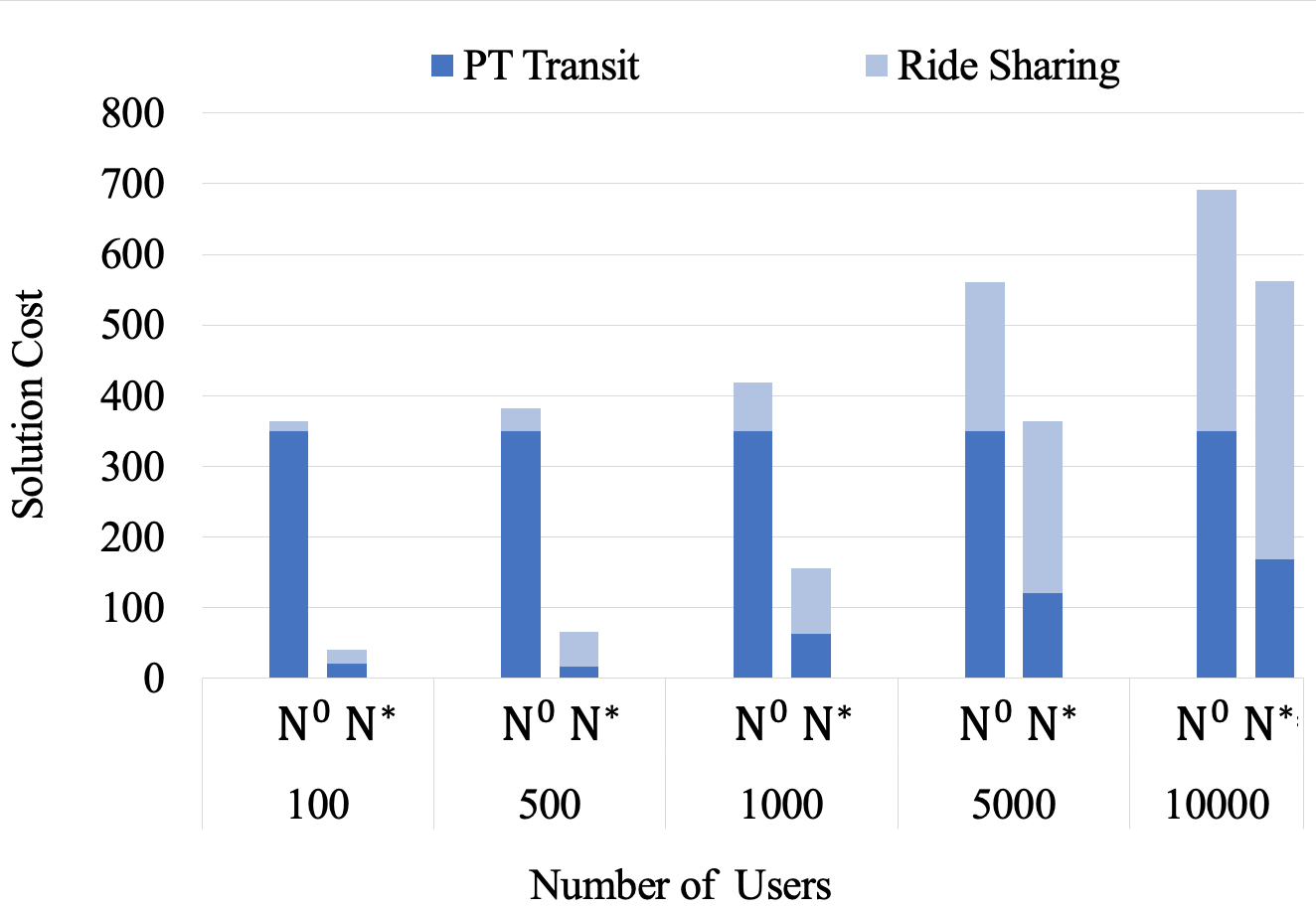}
        \caption{}
        \label{fig:demand}
    \end{subfigure}
    \hfill 
    \begin{subfigure}{.32\textwidth}
        \centering
        \includegraphics[width=\linewidth]{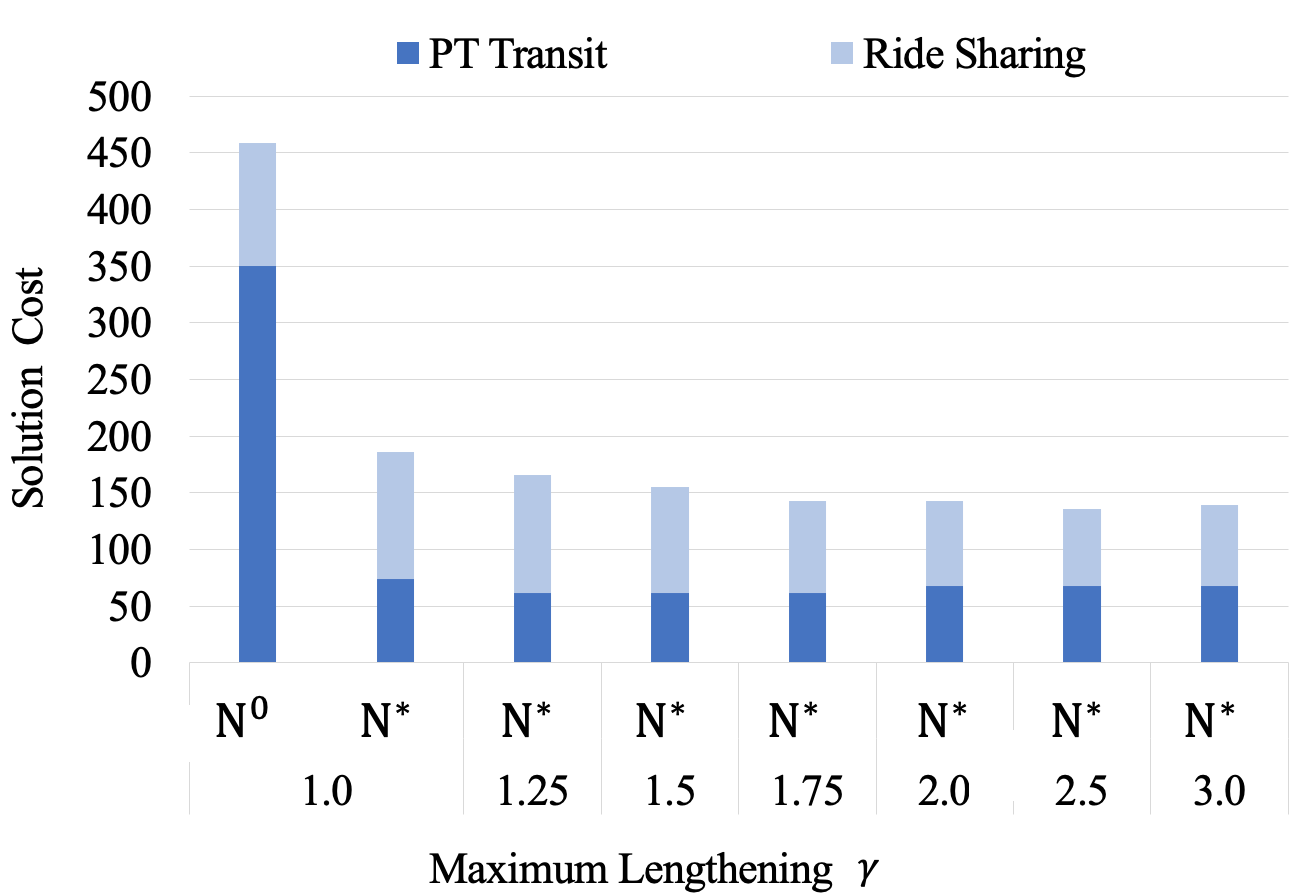}
        \caption{}
        \label{fig:cost}
    \end{subfigure}
    \hfill 
    \begin{subfigure}{.32\textwidth}
        \centering
        \includegraphics[width=\linewidth]{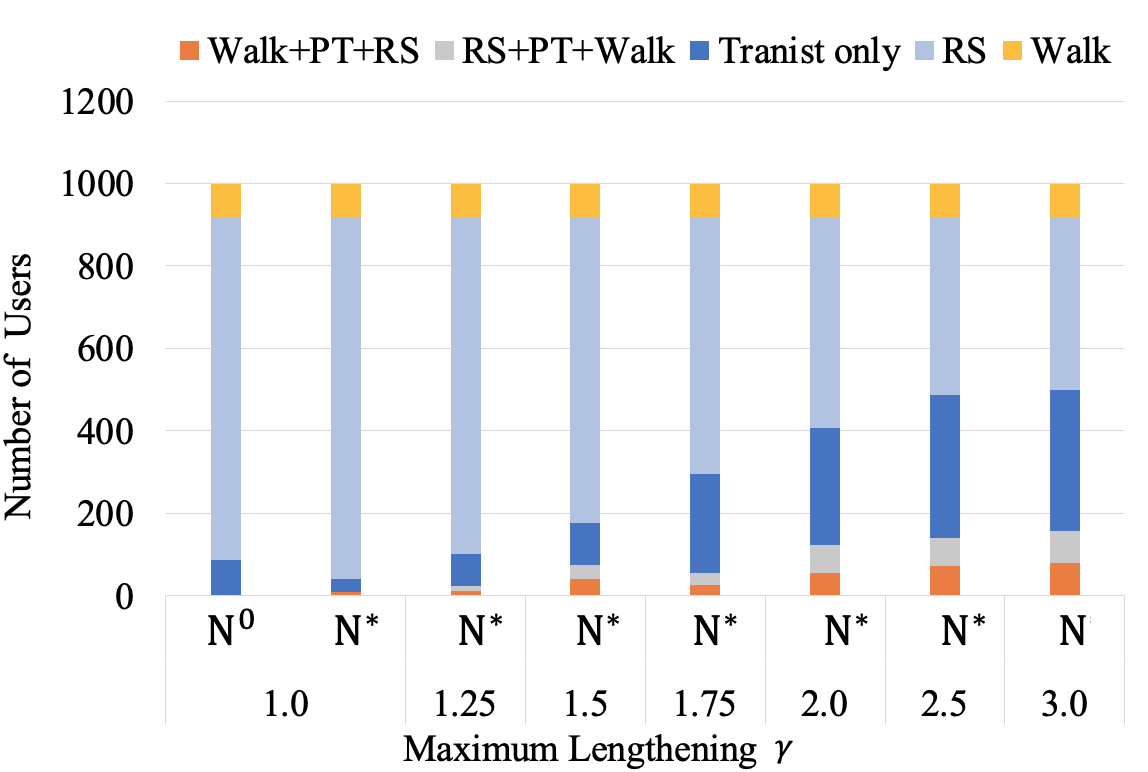}
        \caption{}
        \label{fig:mode}
    \end{subfigure}
    \caption{Performance of the initial layout~$\mathbf{y}^0$ and optimized layout~$\mathbf{y}^*$, under different scenarios.}
    \label{fig:comparative-analysis}
\end{figure}

The optimized layout in the default scenario (with parameters in Table~\ref{tab:parameters}) is shown in Table~\ref{tab:PT_layout_under_default_scenarios}: buses and stops of conventional PT are considerably reduced (and so the operational cost - Fig.~\ref{fig:demand}). Line 7 is completely removed.
%
%
Fig.~\ref{fig:demand} shows that cost reduction is consistent across different levels of demand. However, with few users there is more margin for cost reduction, as conventional PT becomes inefficient and can be well replaced by a relatively small fleet of vehicles.
Fig.~\ref{fig:cost} and~\ref{fig:mode} show that the higher the maximum travel times~$M_i$ tolerated by users (i.e., larger~$\gamma$ - Table~\ref{tab:parameters}), the less RS is used in favor of PT.


\section{Conclusion}
We present a modeling and metaheuristic-based approach to design a multimodal PT, composed of conventional PT and Ride Sharing (RS). We show in simulation that the PT designs issued by our approach can reduce operational cost while respecting users' time constraints. In the future work we will apply this method to a detailed representation of a metropolitan area (current simulations are on a simplified version of the Paris Region) and increase even more the number of considered users to find up to which demand density it is convenient to integrate RS into conventional PT.

\section*{Acknowledgement}
This study is supported by the Research Foundation - Flanders, Belgium (FWO junior research project G020222N).




\begin{thebibliography}{}
\bibitem[Wang et al. (2024)]{Duo2024} Wang, Duo, Andrea Araldo, and Mounim A. El Yacoubi. ``AccEq-DRT: Planning Demand-Responsive Transit to reduce inequality of accessibility.'' Transportation Research Board (TRB) 103rd Annual Meeting. 2024.

\bibitem[Calabro et al.(2023)]{Calabro2023} Calabro, Giovanni, Araldo, Andrea, Oh, Simon, Seshadri, Ravi, Inturri, Giuseppe, Ben-Akiva, Moshe, 2023. "Adaptive transit design: Optimizing fixed and demand responsive multi-modal transportation via continuous approximation." Transportation Research Part A: Policy and Practice, 171, 103643, Elsevier.

\bibitem[Posada et al. (2017)]{Posada2017}Posada, Marcus, Henrik Andersson, and Carl H. Häll. "The integrated dial-a-ride problem with timetabled fixed route service." Public Transport 9 (2017): 217-241.


\bibitem[Molenbruch et al.(2020)]{Mol2020} Molenbruch, Y., Braekers, K., Hirsch, P., Oberscheider, M., 2020. Analyzing the benefits of an integrated mobility system using a matheuristic routing algorithm. European Journal of Operational Research. DOI: 10.1016/j.ejor.2020.07.060.

\bibitem[Stiglic et al.(2018)]{stiglic2018enhancing} Stiglic, Mitja, Agatz, Niels, Savelsbergh, Martin, Gradisar, Mirko, 2018. "Enhancing urban mobility: Integrating ride-sharing and public transit." Computers \& Operations Research, 90, 12--21, Elsevier.

\bibitem[Sun et al.(2018)]{Sun2018} Sun, Bo, Wei, Ming, Yang, Chunfeng, Xu, Zhihuo, Wang, Han, 2018. "Personalised and Coordinated Demand-Responsive Feeder Transit Service Design: A Genetic Algorithms Approach." Future Internet, 10(7), Article 61. DOI: 10.3390/fi10070061.

\bibitem[Posada et al.(2017)]{Pos2017} Posada, M., Andersson, H., 2017. The integrated dial-a-ride problem with timetabled fixed route service. Public Transport 9(1-2), 217-241. DOI: 10.1007/s12469-016-0128-9.

\bibitem[Lee(2017)]{lee2017dynamics} Lee, Cassey, 2017. Dynamics of ride sharing competition. ISEAS Yusof Ishak Institute.

\bibitem[Chow et al.(2019)]{Chow2019} Chow, T.-Y. Ma, S. Rasulkhani, J. Y.J. Chow, S. Klein, 2019. A dynamic ridesharing dispatch and idle vehicle repositioning strategy with integrated transit transfers. Transportation Research Part E: Logistics and Transportation Review 128, 417-442. DOI: 10.1016/j.tre.2019.07.002.


\bibitem[Boeing(2019)]{Boeing2019}
Boeing, Geoff, 2019. Urban spatial order: street network orientation, configuration, and entropy. Applied Network Science, 4(1). DOI: 10.1007/s41109-019-0189-1.


\bibitem[Cascetta (2009)]{Cascetta2009} Cascetta, E. (2009). Transportation systems analysis: models and applications. Springer Science \& Business Media.

\bibitem[Khanesar et al.(2007)]{Kha2007}
Khanesar, Mojtaba Ahmadieh and Teshnehlab, Mohammad and Aliyari Shoorehdeli, Mahdi, 2007. A novel binary particle swarm optimization. 2007 Mediterranean Conference on Control \& Automation, 1-6.

\bibitem[Cipriani et al.(2020)]{cipriani2020}
Cipriani, E., Fusco, G., Patella, S. M., \& Petrelli, M., 2020. A Particle Swarm Optimization Algorithm for the Solution of the Transit Network Design Problem. Smart Cities, 3(2), 541--555. https://doi.org/10.3390/smartcities3020029

\bibitem[Yong-chuan et al.(2011)]{yong2011traffic}
Yong-chuan, Zhang and Xiao-qing, Zuo and Zhen-ting, Chen and others, 2011. Traffic congestion detection based on GPS floating-car data. Procedia Engineering 15, 5541--5546.

\bibitem[Domínguez et al.(2014)]{dominguez2014multi}
Domínguez, María and Fernández-Cardador, Antonio and Cucala, Asunción P and Gonsalves, Tad and Fernández, Adrián, 2014. Multi objective particle swarm optimization algorithm for the design of efficient ATO speed profiles in metro lines. Engineering Applications of Artificial Intelligence 29, 43--53.

\bibitem[Giacomin and Levinson(2015)]{Levinson2015}
Giacomin, David J and Levinson, David M, 2015. Road network circuity in metropolitan areas. Environment and Planning B: Planning and Design 42.6, 1040--1053.

\bibitem[Zhao and Deng(2013)]{zhao2013relationship}
Zhao, Jinbao and Deng, Wei, 2013. Relationship of walk access distance to rapid rail transit stations with personal characteristics and station context. Journal of urban planning and development 139.4, 311--321.

\bibitem[Sadrani et al.(2022)]{AntoniouSadrani2022}
Sadrani, Mohammad and Tirachini, Alejandro and Antoniou, Constantinos, 2022. Vehicle dispatching plan for minimizing passenger waiting time in a corridor with buses of different sizes: Model formulation and solution approaches. European Journal of Operational Research 299.1, 263--282.

\bibitem[Bosch et al.(2018)]{bosch2018cost} Bösch, P. M., Becker, F., Becker, H., Axhausen, K. W., 2018. Cost-based analysis of autonomous mobility services. Transport Policy 64, 76-91. DOI: 10.1016/j.tranpol.2017.09.005.


\bibitem[Cats et al.(2021)]{CatsJointOptim2021} Zhao, Jing, Sun, Sicheng, Cats, Oded, 2021. "Joint optimisation of regular and demand-responsive transit services." Transportmetrica A: Transport Science, https://doi.org/10.1080/23249935.2021.198758.

\bibitem[Steiner and Irnich(2020)]{steiner2020strategic}Steiner, K., Irnich, S., 2020. Strategic planning for integrated mobility-on-demand and urban public bus networks. Transportation Science 54(6), 1616--1639. Publisher: INFORMS.


\bibitem[Zhao and Deng(2013)]{zhao2013relationship}
Zhao, Jinbao and Deng, Wei, 2013. Relationship of walk access distance to rapid rail transit stations with personal characteristics and station context. Journal of urban planning and development 139.4, 311--321.

\bibitem[Bosch et al.(2018)]{bosch2018cost} Bösch, P. M., Becker, F., Becker, H., Axhausen, K. W., 2018. Cost-based analysis of autonomous mobility services. Transport Policy 64, 76-91. DOI: 10.1016/j.tranpol.2017.09.005.


\bibitem[Omnil (2019)]{IdF2019} Omnil-Ile de France Mobilités. La nouvelle enquête globale transport Présentation des premiers resultats 2018. Technical report, 2019.

\end{thebibliography}
\end{document}